\DocumentMetadata{}
\documentclass[sigplan,nonacm]{acmart}

\usepackage{tikz}
\usepackage{xspace}
\usepackage{siunitx}
\usepackage{amsmath}
\usepackage{subfig}
\usepackage{soul}

\usepackage{hyperref}
\usepackage{listings}

\definecolor{codegreen}{rgb}{0,0.6,0}
\definecolor{codegray}{rgb}{0.5,0.5,0.5}
\definecolor{codepurple}{rgb}{0.58,0,0.82}
\definecolor{backcolour}{rgb}{1, 1, 1}

\lstdefinestyle{mystyle}{
  backgroundcolor=\color{backcolour}, commentstyle=\color{codegreen},
  keywordstyle=\color{magenta},
  numberstyle=\tiny\color{codegray},
  stringstyle=\color{codepurple},
  basicstyle=\ttfamily\footnotesize,
  breakatwhitespace=false,         
  breaklines=true,                 
  captionpos=b,                    
  keepspaces=true,                 
  numbers=left,                    
  numbersep=5pt,                  
  showspaces=false,
  showstringspaces=false,
  showtabs=false,                  
  tabsize=2
}

\newcommand*\circled[2]{\tikz[baseline=(char.base)]{
            \node[shape=circle,fill=black,inner sep=1pt] (char) {\textcolor{#1}{{\footnotesize #2}}};}}

\ifx\figurename\undefined \def\figurename{Figure}\fi
\renewcommand{\figurename}{Fig.}
\renewcommand{\paragraph}[1]{\textbf{#1} }

\newcommand{\Sect}[1]{Sec.~\ref{#1}}
\newcommand{\Fig}[1]{Fig.~\ref{#1}}
\newcommand{\Tbl}[1]{Tbl.~\ref{#1}}

\newcommand{\Eqn}[1]{Eqn.~\ref{#1}}

\newcommand{\specialcell}[2][c]{\begin{tabular}[#1]{@{}c@{}}#2\end{tabular}}

\def\cC{{\mathcal{C}}}
\def\cF{{\mathcal{F}}}
\def\cM{{\mathcal{M}}}
\def\cS{{\mathcal{S}}}
\def\cR{{\mathcal{R}}}

\newcommand{\proj}{\textsc{StreamGrid}\xspace}

\newcommand{\mode}[1]{\underline{\textsc{#1}}\xspace}

\newcommand{\RNum}[1]{\uppercase\expandafter{\romannumeral #1\relax}}

\graphicspath{{figs/}}

\copyrightyear{2025}
\acmYear{2025}
\setcopyright{acmlicensed}\acmConference[ASPLOS '25]{Proceedings of the 30th ACM International Conference on Architectural Support for Programming Languages and Operating Systems, Volume 2}{March 30-April 3, 2025}{Rotterdam, Netherlands}
\acmBooktitle{Proceedings of the 30th ACM International Conference on Architectural Support for Programming Languages and Operating Systems, Volume 2 (ASPLOS '25), March 30-April 3, 2025, Rotterdam, Netherlands}
\acmDOI{10.1145/3676641.3716021}
\acmISBN{979-8-4007-1079-7/2025/03}

\makeatletter
\gdef\@copyrightpermission{
  \begin{minipage}{0.3\columnwidth}
       \href{https://creativecommons.org/licenses/by/4.0/}{%
      \includegraphics[width=0.90\textwidth]{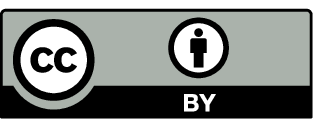}%
    }
  \end{minipage}\hfill
  \begin{minipage}{0.7\columnwidth}
       \href{https://creativecommons.org/licenses/by/4.0/}{%
      This work is licensed under a Creative Commons Attribution International 4.0 License.%
    }
  \end{minipage}
  \vspace{5pt}
}
\makeatother

\begin{document}

\title[\proj]{\proj: Streaming Point Cloud Analytics via Compulsory Splitting and Deterministic Termination}



\author{Yu Feng}
\authornote{All authors contributed equally to this research.}
\authornote{Corresponding Authors.}
\orcid{0000-0002-2192-5737}
\affiliation{%
  \institution{Shanghai Jiao Tong University, Shanghai Qi Zhi Institute}
  \city{Shanghai}
  \country{China}
}
\email{y-feng@sjtu.edu.cn}

\author{Zheng Liu}
\authornotemark[1]
\orcid{0009-0001-6688-4115}
\affiliation{%
  \institution{Shanghai Jiao Tong University}
  \city{Shanghai}
  \country{China}
}
\email{distilledw@sjtu.edu.cn}

\author{Weikai Lin}
\authornotemark[1]
\orcid{0000-0003-3537-4857}
\affiliation{%
  \institution{University of Rochester}
  \city{Rochester}
  \state{NY}
  \country{USA}
}
\email{wlin33@ur.rochester.edu}

\author{Zihan Liu}
\orcid{0000-0002-0874-0682}
\affiliation{%
  \institution{Shanghai Jiao Tong University, Shanghai Qi Zhi Institute}
  \city{Shanghai}
  \country{China}
}
\email{altair.liu@sjtu.edu.cn}

\author{Jingwen Leng}
\authornotemark[2]
\orcid{0000-0002-5660-5493}
\affiliation{%
  \institution{Shanghai Jiao Tong University, Shanghai Qi Zhi Institute}
  \city{Shanghai}
  \country{China}
}
\email{leng-jw@sjtu.edu.cn}

\author{Minyi Guo}
\orcid{0000-0003-0034-2302}
\affiliation{%
  \institution{Shanghai Jiao Tong University, Shanghai Qi Zhi Institute}
  \city{Shanghai}
  \country{China}
}
\email{guo-my@sjtu.edu.cn}

\author{Zhezhi He}
\orcid{0000-0002-6357-236X}
\affiliation{%
  \institution{Shanghai Jiao Tong University}
  \city{Shanghai}
  \country{China}
}
\email{zhezhi.he@sjtu.edu.cn}

\author{Jieru Zhao}
\orcid{0000-0001-8211-2812}
\affiliation{%
  \institution{Shanghai Jiao Tong University}
  \city{Shanghai}
  \country{China}
}
\email{zhao-jieru@sjtu.edu.cn}

\author{Yuhao Zhu}
\orcid{0000-0002-2802-0578}
\affiliation{%
  \institution{University of Rochester}
  \city{Rochester}
  \state{NY}
  \country{USA}
}
\email{yzhu@rochester.edu}

\renewcommand{\shortauthors}{}

\keywords{Hardware Accelerator, Streaming Architecture, Line Buffer, Point Cloud Analytics.}



\begin{abstract}

Point clouds are increasingly important in intelligent applications, but frequent off-chip memory traffic in accelerators causes pipeline stalls and leads to high energy consumption. 
While conventional line buffer techniques can eliminate off-chip traffic, they cannot be directly applied to point clouds due to their inherent computation patterns. 
To address this, we introduce two techniques: \textit{compulsory splitting} and \textit{deterministic termination}, enabling fully-streaming processing. 
We further propose \proj, a framework that integrates these techniques and automatically optimizes on-chip buffer sizes. 
Our evaluation shows \proj reduces on-chip memory by 61.3\% and energy consumption by 40.5\% with marginal accuracy loss compared to the baselines without our techniques.
Additionally, we achieve 10.0$\times$ speedup and 3.9$\times$ energy efficiency over state-of-the-art accelerators.

\end{abstract}
\hypersetup{bookmarks=true,breaklinks=true,letterpaper=true,colorlinks,linkcolor=blue,citecolor=magenta,urlcolor=blue}

\maketitle 

\section{Introduction}
\label{sec:intro}

In recent years, point clouds have emerged as one of the primary modalities in intelligent applications~\cite{guo2020deep, pomerleau2015review}. 
Their versatility has enabled widespread adoption across numerous domains, including autonomous driving~\cite{cui2021deep, feng2020real}, robotics~\cite{rusu2008towards}, 3D modeling~\cite{kerbl20233d}, and augmented reality~\cite{stets2017visualization}.
These algorithmic advancements have also spurred the development of various domain-specific accelerators in the hardware community to improve the efficiency of point cloud processing~\cite{zhang2021point, lin2021pointacc, feng2020mesorasi, feng2022crescent, xu2019tigris, ying2023edgepc, han2024bitnn}.

Despite significant progress in point cloud accelerators, one fundamental issue remains unaddressed: frequent off-chip memory traffic. 
Unlike image signal processing, where line-buffered accelerators~\cite{ujjainkar2023imagen, chi2018soda,whatmough2019fixynn, hegarty2014darkroom} effectively store intermediate results and minimize off-chip traffic, point cloud accelerator designs often rely on double buffering to possibly hide off-chip latency when reading/writing the intermediate data between stages~\cite{zhang2021point, lin2021pointacc, feng2020mesorasi}.
Although this approach can have performance guarantees, it leads to substantial energy consumption, particularly for mobile devices, as the off-chip memory access energy is orders of magnitude higher than the on-chip one~\cite{gao2017tetris, Yazdanbakhsh2018GAN}.

Line buffer techniques work well in image signal processing because these applications primarily involve stencil operations.
Each output element often relies on nearby data and exhibits fixed processing patterns, requiring only minimal buffering between stages.

In contrast, operations in point cloud applications are more diverse.
They can be classified into two main categories: those with local data dependencies and those with global data dependencies (\Sect{sec:bg}).
Global data dependencies imply that a single output point may need to access all input data points. 
Due to those operations with global dependencies, applying line buffer techniques directly to point cloud applications would result in unaffordable on-chip buffer sizes, as the worst-case scenario would require buffering all intermediate data on-chip. 

Additionally, line buffer techniques require deterministic throughput and memory access patterns for each stage.
However, most operations with global dependencies are non-deterministic in point cloud applications. 
This means that their throughput is input-dependent, making it impossible to determine the optimal line buffer size offline.

\paragraph{Main Ideas.} 
We propose two principle techniques to address these issues and evaluate them across four application domains to ensure minimal impact on application accuracy (\Sect{sec:algo}). 
First, we introduce \textit{compulsory splitting}, which partitions the original point cloud data into smaller chunks and intentionally relaxes some data dependencies across chunks, allowing each chunk to be processed independently. 
This approach significantly reduces on-chip buffer requirements and enables finer-grained pipelining across multiple chunks. 

Second, we propose \textit{deterministic termination}, which converts non-deterministic operations into deterministic ones by setting a fixed termination “deadline” for each operation.
This forces operations to terminate after a predefined number of steps, ensuring that the delays of non-deterministic operations are no longer input-dependent.

Coupled with these two techniques, we propose an integrated co-training procedure that incorporates our proposed algorithmic behaviors into the training process.
This ensures that co-trained models remain robust against these algorithmic modifications with only marginal accuracy loss.

\begin{figure*}[t]
\centering
\includegraphics[width=\textwidth]{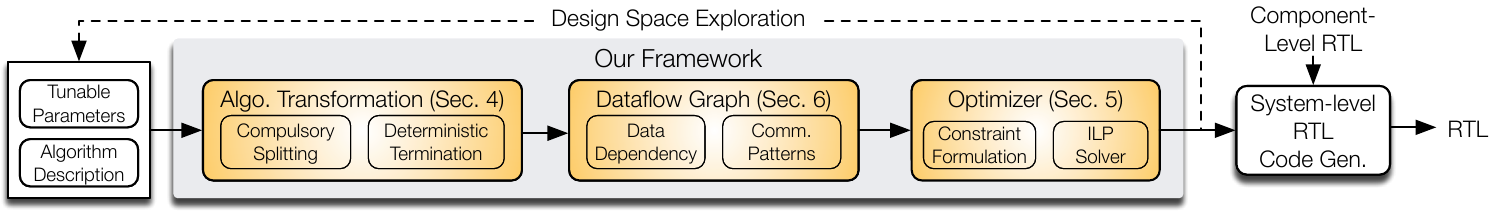}
\caption{Our framework first takes the original algorithm description with tuneable parameters and performs the algorithmic optimizations (\Sect{sec:algo}). The transformed algorithm is then used to form a dataflow graph and perform dependency and data reuse analyses (\Sect{sec:framework}). Lastly, this graph forms a set of constraints and optimizes the optimal line buffer size using ILP (\Sect{sec:opti}).}
\label{fig:workflow}
\end{figure*}

\paragraph{Framework.} 
The two techniques above relax the global dependencies in point cloud algorithms and convert non-deterministic point cloud pipelines into deterministic ones, enabling pipelining across different stages.
The remaining challenge is to determine the minimal line buffer sizes between stages.
To address this, we propose \proj, a unified framework for point cloud applications, which automatically searches for optimal line buffer sizes (\Fig{fig:workflow}).

To accommodate a wide range of point cloud operations, we provide an abstract interface that allows users to describe dataflow graphs of the algorithms without specifying the operation details (\Sect{sec:framework}). 
We then extract data dependencies and communication patterns from the users’ dataflow graphs and form a constrained line buffer minimization problem. 
Using an integer linear programming (ILP) solver, we can solve this minimization problem and identify the optimal on-chip buffer size under a given performance target (\Sect{sec:opti}).

One key challenge in this ILP formulation is that the search space quickly becomes unmanageable as the number of optimization parameters increases. 
To address this, we exploit the monotonicity in line buffer constraints and propose a constraint pruning method that drastically reduces the number of optimization parameters.

\paragraph{Result.} Compared to the baselines without using our techniques, we demonstrate that \proj achieves 61.3\% on-chip memory reduction and 40.5\% energy savings with marginal accuracy loss across four application domains. 
We also evaluate our techniques against five state-of-the-art accelerator designs and show that, on average, our accelerator delivers 10.0 $\times$ speedup and 3.9$\times$ energy reduction, all with comparable hardware resources.

The contributions of this paper are as follows:
\begin{itemize}
\item We propose two techniques, \textit{compulsory splitting} and \textit{deterministic termination}, that reduce on-chip buffer requirements and regularize the non-determinism in point cloud pipelines.
\item We propose \proj, the first framework leveraging line buffer techniques for point cloud, that automatically searches for minimal line buffer sizes.
\item We demonstrate that \proj enables streaming architecture design with marginal accuracy loss across four application domains.
\end{itemize}

\section{Background}
\label{sec:bg}

We first introduce the point cloud applications in \Sect{sec:bg:pc}, and then explain the basics of line buffer design in \Sect{sec:bg:lb}.

\subsection{Point Cloud-based Applications}
\label{sec:bg:pc}


\paragraph{Categories.} 
Overall, point cloud applications can be classified into two main categories: conventional pipelines and deep neural network (DNN)-based pipelines. 
While DNN-based pipelines have revolutionized many fields recently, conventional pipelines often achieve superior performance in areas such as mapping~\cite{whitty2010autonomous}, localization~\cite{liang2018deep}, and more~\cite{pomerleau2015review}. 
These pipelines typically include operations such as filtering, sampling, and nearest neighbor search. 
On the other side, DNN-based pipelines become the predominant methods in tasks like classification, detection, and more~\cite{guo2020deep, liu2019deep, feng2023fast}. 
Recent developments in neural rendering also utilize point clouds as the rendering primitives in novel view synthesis and reconstruction~\cite{kerbl20233d, fang2024mini, fan2023lightgaussian}.
Compared to conventional pipelines, these DNN-based methods often include additional operations like convolutions and multilayer perceptions.

\begin{figure}[t]
\centering
\subfloat[\textit{Local-dependent operations} include, on the left, a reduction operation that computes the maximum along a direction, and on the right, a $1\times3$ stencil operation that computes the curvature of each point using 2 adjacent points.]{
	\label{fig:local_ops}	
        \includegraphics[width=\columnwidth]{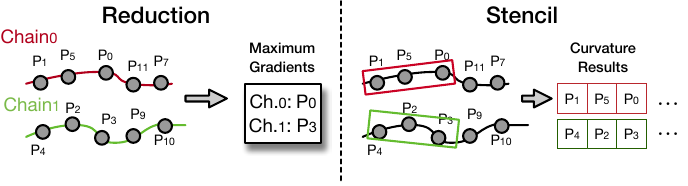} }
\\
\vspace{-5pt}
\subfloat[\textit{Global-dependent operations} include, on the left, a sorting operation that sorts all points along the y-axis, and on the right, a kNN search operation that identifies the $k$ nearest neighbors of a query point (e.g., $P_1$ and $P_2$).
]{
	\label{fig:global_ops}
	\includegraphics[width=\columnwidth]{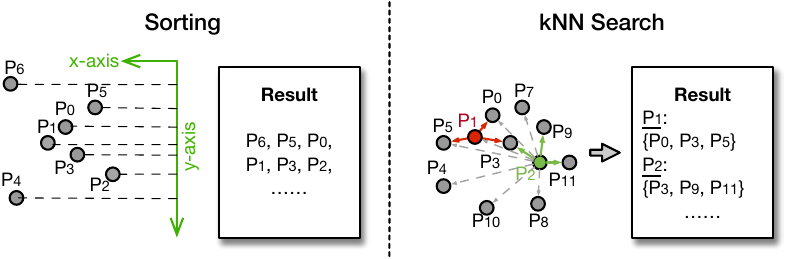} } 
\caption{Point cloud operations, including those with local data dependency and those with global data dependency. 
}
\vspace{-5pt}
\label{fig:ops}
\end{figure}

Unlike image signal processing, where the primary operations are stencil operations~\cite{ragan2013halide, chi2018soda, ujjainkar2023imagen, whatmough2019fixynn, hegarty2014darkroom, mullapudi2016automatically}, point cloud processing tasks involve various operations. 
Based on data dependency, point cloud operations can be classified into two main categories: those with local data dependency and those with global data dependency. 
Global data dependency implies that, algorithmically, one output point potentially requires access to all the input data points, whereas local data dependency does not.
\Fig{fig:ops} shows some examples.

\paragraph{Local Dependency.}
Operations with local data dependency, referred to as \textit{local-dependent operations}, typically involve processing points within close proximity.
Examples include elementwise operations where points are modified independently, such as scaling or thresholding.
Additional examples are operations involving local neighbor points, such as reduction and stencil operations, as shown in \Fig{fig:local_ops}.
On the left, \Fig{fig:local_ops} shows a reduction operation that determines the maximum value among a chain of points. 
On the right, it shows a stencil operation that performs convolution operations in a sliding window fashion with a window size of $1\times3$ and computes the curvature at each point by incorporating two adjacent points.

\paragraph{Global Dependency.}  
However, determining the spatial relationships within a point cloud often involves operations with global data dependencies that span across all points. 
We refer to these operations as \textit{global-dependent operations}.
In today's point cloud applications, the primary operations include sorting, range search, and k-nearest neighbors (kNN) search, as shown in \Tbl{tab:eval_app}.
\Fig{fig:global_ops} gives some examples. 
On the left, it shows a sorting operation that sorts all points along the y-axis.
On the right, \Fig{fig:global_ops} illustrates a kNN search, which involves identifying the $k$ nearest point neighbors within a point cloud using data structures like kd-trees~\cite{bentley1975multidimensional, zhou2008real}. 
Note that, although operations like kNN search only need to identify nearby points, the search process itself may require iterating all points to obtain results as the dashed lines show. 

\subsection{Line Buffer}
\label{sec:bg:lb}

\begin{figure}[t]
\centering
\includegraphics[width=\columnwidth]{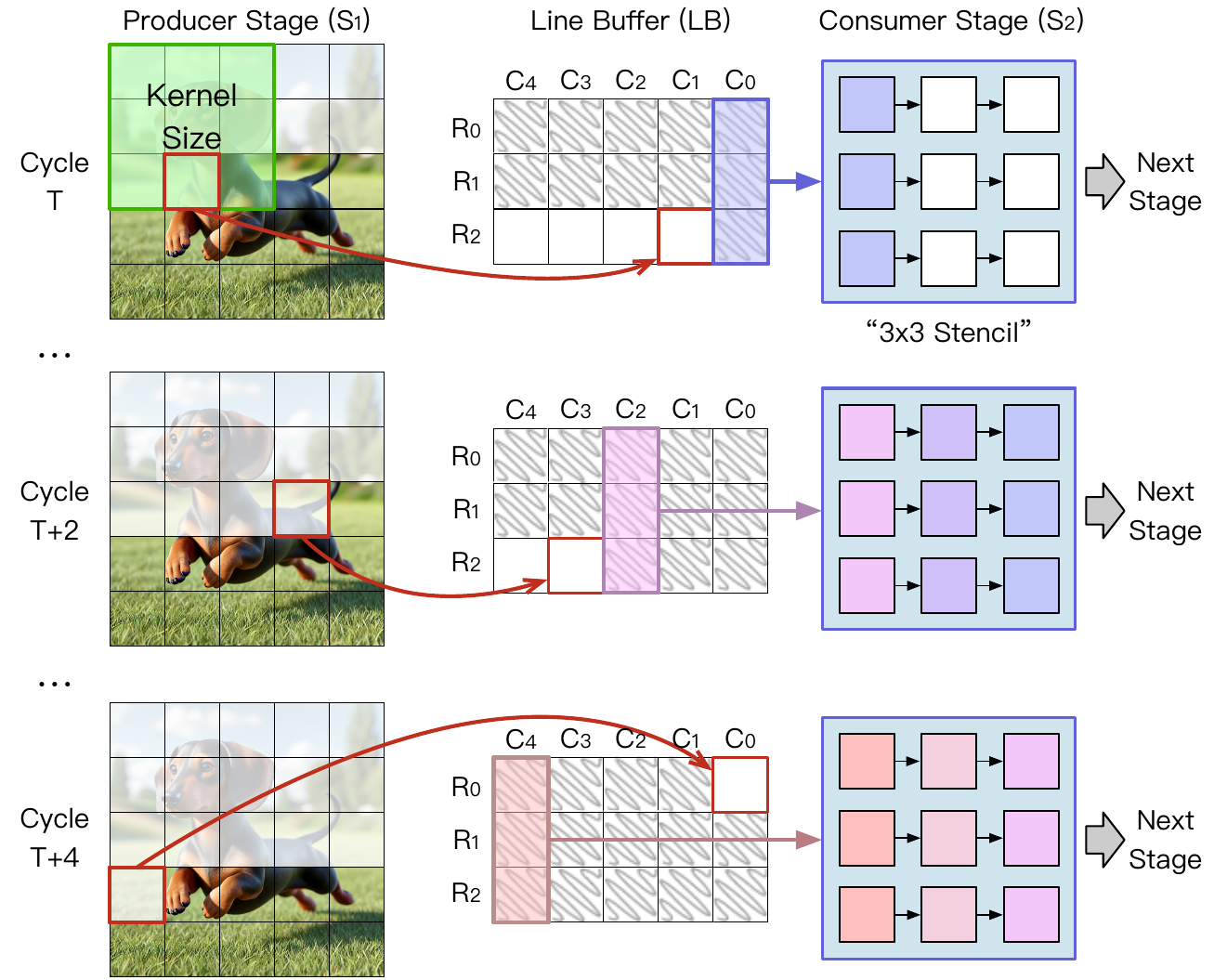}
\caption{An example of using a $3\times5$ line buffer with a $3\times3$ stencil operation in an image processing pipeline.}
\label{fig:line_buffer_example}
\end{figure}

Line buffers play a crucial role in image processing accelerators, where the target pipelines primarily consist of stencil operations~\cite{ragan2013halide, chi2018soda, ujjainkar2023imagen, whatmough2019fixynn, hegarty2014darkroom, mullapudi2016automatically}. 
In these operations, the new value of each pixel is often calculated by applying a kernel to a neighborhood of pixels, as shown in \Fig{fig:line_buffer_example}.
In this case, the kernel size is $3\times3$ highlighted in green. 
This characteristic enables image processing accelerators to leverage line buffers, temporarily storing the necessary rows or columns of pixels from the producer stage.
This allows the consumer stage to access and compute the values of neighboring pixels directly from the line buffer, enabling a streaming process that eliminates off-chip traffic.

For instance, in \Fig{fig:line_buffer_example}, a ``dachshund'' image undergoes a $3\times3$ stencil operation. 
At each cycle, the producer stage, $S_1$, generates and writes one pixel into the line buffer, $LB$, as highlighted by the red bounding box.

From cycle $0$, $S_1$ continues writing pixels into $LB$ in row-major order, and the consumer stage, $S_2$, stays idle.
Until cycle $T$, the pixel is written into $LB$ at position $\langle R_2, C_1 \rangle$. 
At the same cycle, $S_2$ begins reading pixels from column $C_0$ of $LB$. From this point forward, in each subsequent cycle, $S_2$ continues to read a column of pixels from $LB$ until the entire stencil computation is completed.

However, $S_2$ does not output results until cycle $T+2$. At cycle $T+2$, $S_2$ has gathered all the required pixels for the stencil operation (in $3 \times 3$ shift registers). It performs a convolution operation and writes the first output pixel to the next processing stage.
Note that, the line buffer, $LB$, is designed only large enough to accommodate three rows of pixels. When $S_1$ outputs the first element from the fourth row, it overwrites the line buffer at position $\langle R_0, C_0 \rangle$. This overwrite occurs because, from cycle $T+4$, the original pixel at $\langle R_0, C_0 \rangle$ is no longer needed for future computations and can be safely discarded.

To sum up, storing intermediate values in line buffers eliminates the need to write data back to off-chip memory after each operation. 
However, applying this technique requires:
\begin{itemize}
    \item Operations should ideally be local-dependent so that all dependent data can be accommodated on-chip.
    \item Both the producer and consumer must have fixed throughputs so that the minimal line buffer size can be pre-determined offline.
    \item The data access pattern of each stage must be regular, ensuring that the read/write ports required for each line buffer can be determined without bank conflicts.
\end{itemize}

\section{Challenges}
\label{sec:ch}



Applying line buffers in point cloud applications presents unique challenges other than their use in image processing. 

\paragraph{Global-Dependent Operations.} 
The first challenge is that \textit{global-dependent operations would lead to unaffordable on-chip buffers and disrupt fine-grained pipelining}. 

Unlike image processing where data dependencies are localized, these point cloud operations potentially require to access all points in a point cloud.
As a result, the line buffer can become too large, making it infeasible for mobile devices. 
For instance, sorting operations require accessing all the data to establish the correct order. 
The required buffer size in sorting is proportional to the point cloud size.
Sorting a point cloud with half a million points using bitonic sort requires buffering over 30 million elements, i.e., 30 MB of on-chip buffer, which is impractical for mobile SoCs~\cite{xaviersoc, orinsoc, snapdragonxr2}.


Another benefit of using line buffers is to achieve fine-grained pipelining across stages, allowing the execution of different stages to overlap. 
However, global-dependent operations disrupt the pipeline due to their data dependencies, i.e., the global-dependent consumer must wait until all the data from its producer are ready before it can proceed.

\begin{figure}[t]
\centering
\includegraphics[width=\columnwidth]{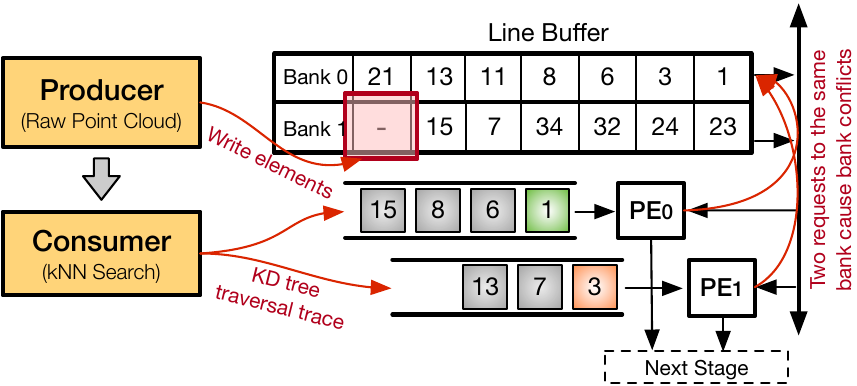}
\caption{An example of non-deterministic operations in point cloud applications: 1) the throughput of each PE is input-dependent; 2) irregular memory access patterns can lead to bank conflicts and pipeline stalls.}
\label{fig:non_deterministic_example}
\end{figure}

\paragraph{Non-Determinism.} 
The second challenge is that \textit{many point cloud operations are non-deterministic by nature.}  
This non-determinism leads to two key issues.
First, the latency of a non-deterministic operation is input-dependent, resulting in variable throughput. 
This makes it difficult to determine the optimal size for line buffers, as the required buffer size between a producer and a consumer can change dynamically based on the inputs.
Second, non-deterministic operations often exhibit irregular memory access patterns, which cause bank conflicts, potentially stalling the entire pipeline.

\Fig{fig:non_deterministic_example} gives an example of a kNN search using a kd-tree.
The producer writes the raw point cloud data into a line buffer, and the consumer has two processing elements (PEs) that perform kd-tree searches in parallel. 
In \Fig{fig:non_deterministic_example}, each PE follows a different tree traversal path. 
In this case,  $PE_0$  needs to traverse 4 nodes, while  $PE_1$  needs to traverse 3 nodes, leading to variation in execution steps.
We profile the distribution of the number of steps in a kd-tree traversal for a point cloud in the KITTI dataset~\cite{Geiger2013kitti}, where each point searches for 32 neighbors. 
The result shows the average number of steps is $8.4\times10^3$ with a large standard deviation of $6.8\times10^3$ across all points.

Additionally, \Fig{fig:non_deterministic_example} also shows the potential bank conflicts that could happen as the data accesses in neighbor search are input-dependent. 
In this case, both $PE_0$ and $PE_1$ access different elements ($1$ and $3$) in bank 0, causing one of the accesses to be stalled.

To summarize, directly applying line buffers in point cloud pipelines has two main issues. 
First, global-dependent operations lead to unaffordable on-chip buffers and disrupt fine-grained pipelining.
Second, the delays and memory accesses of some operations are input-dependent, resulting in unavoidable pipeline stalls.


\section{Fully-Streaming Point Cloud Processing}
\label{sec:algo}

This section introduces two techniques to address the challenges in \Sect{sec:ch}. 
We first describe \textit{compulsory splitting} (\Sect{sec:algo:ds}) to address the issues with global-dependent operations, we then describe \textit{deterministic termination} (\Sect{sec:algo:dt}) to address the challenges in non-deterministic operations. 
Lastly, we propose a training method that integrates both techniques and guarantees marginal accuracy loss (\Sect{sec:algo:train}).


\begin{figure}[t]
\centering
\begin{minipage}[t]{0.47\columnwidth}
  \centering
  \includegraphics[width=0.93\columnwidth]{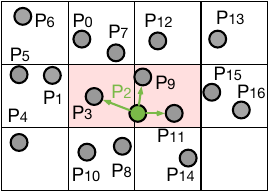}
  \caption{One query point only accesses a limited number of chunks. Here, $P_2$ only accesses 2 highlighted chunks.}
  \label{fig:chunk_example}
\end{minipage}
\hspace{2pt}
\begin{minipage}[t]{0.49\columnwidth}
  \centering
  \includegraphics[width=\columnwidth]{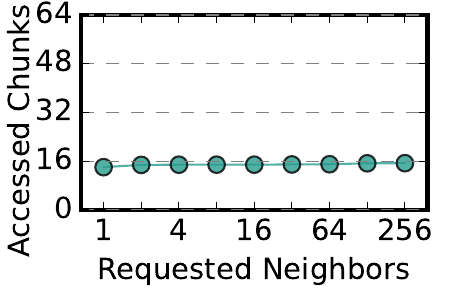}
  \caption{The average number of accessed chunks vs. the number of requested neighbors per point.}
  \label{fig:block_access}
\end{minipage}
\end{figure}

\subsection{Compulsory Splitting}
\label{sec:algo:ds}

\paragraph{Observation.} 
Algorithmically, each output element in a global-dependent operation has data dependencies across all input elements.
However, in practice, these elements often interact with only a small subset of the data.
This is because global-dependent operations, such as kd-tree search or hierarchical sorting, inherently perform spatial partitioning to establish local relationships between points.
As a result, these operations exhibit ``local dependencies'' during actual execution.
For instance, in kNN search, 
if we partition the entire point cloud into $4 \times 3$ spatially even chunks as shown in \Fig{fig:chunk_example}, a query point ($P_2$) searching for 3 neighbors would access only a limited subset of these chunks (2 in this case) during the search process.

\Fig{fig:block_access} shows the correlation between the average number of accessed chunks and the number of requested neighbors in the kNN search on the KITTI dataset. 
We partition the input point cloud into $8\times8$ chunks.
For each query point, as the number of neighbors increases, the number of accessed chunks also rises.
However, even when requesting up to 256 neighbors, the average number of accessed chunks remains low, on average 16.
This shows that even global-dependent operations only interact with a small region of data. 

\paragraph{Idea.} 
Leveraging this key insight, our idea is to partition the original input data into smaller chunks, relaxing the data dependency across chunks.
In this way, we can reduce the line buffer sizes for global-dependent operations.
We call this technique, \textit{compulsory splitting}.

\begin{figure}[t]
\centering
\includegraphics[width=\columnwidth]{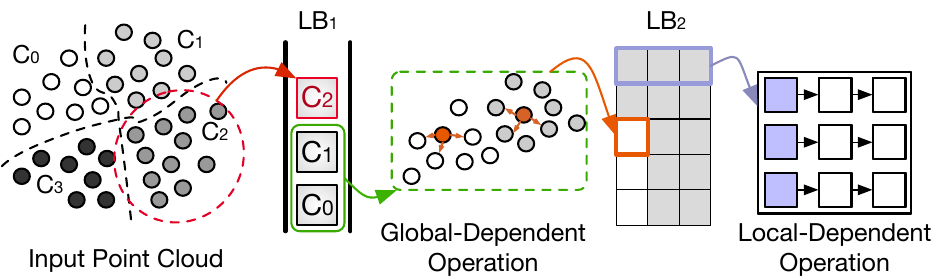}
\caption{An example of applying compulsory splitting to a global-dependent operation. We first partition the point clouds into small chunks and feed each chunk into a line buffer. The global-dependent operation (e.g., kNN search) reads the chunks in a $1\times2$ sliding window fashion and performs a kNN search for every two chunks. 
}
\label{fig:split_lb}
\end{figure}

\begin{figure}[t]
\centering
\includegraphics[width=\columnwidth]{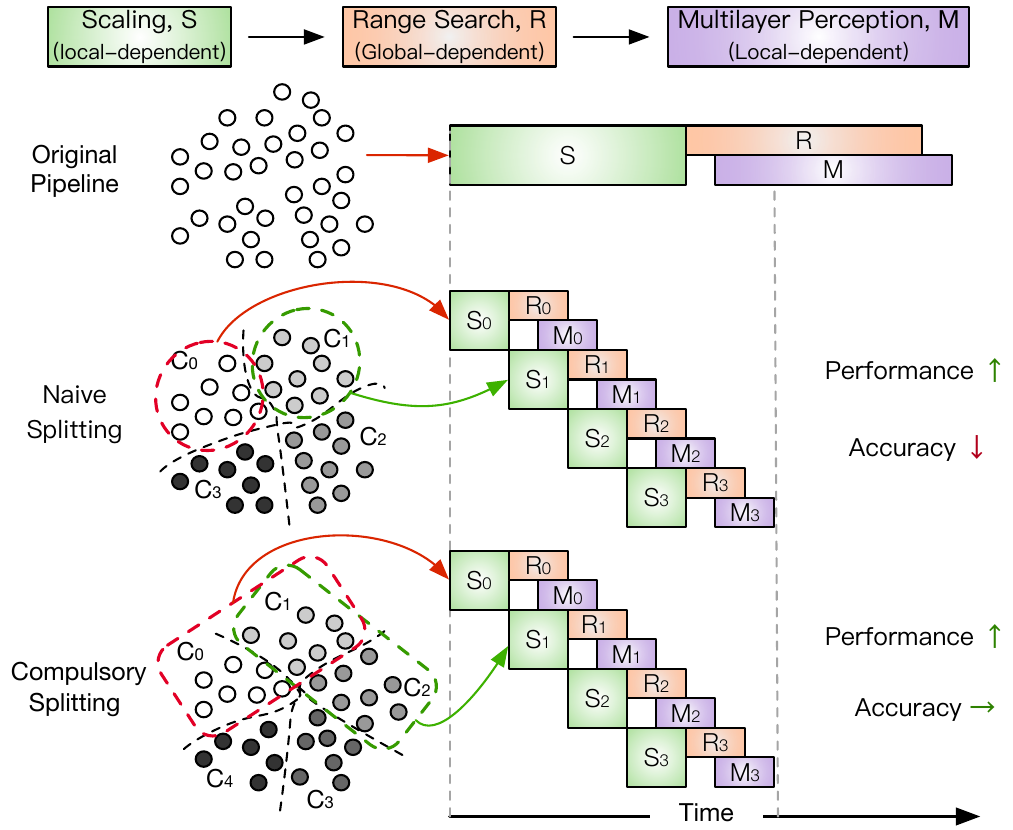}
\caption{An example of applying compulsory splitting. By splitting a point cloud into multiple chunks, we can unlock more fine-grained parallelism. Compared to naive splitting, our technique can preserve some data dependency across neighboring chunks, avoiding accuracy loss. 
}
\label{fig:split_example}
\end{figure}

A naive splitting is to split the original point cloud into completely independent chunks and process them separately. 
For example, the input point cloud is divided into 4 chunks and processed individually in \Fig{fig:split_example}.
By doing so, we can reduce the required line buffer sizes for global-dependent operations and have a finer-grained pipeline.
However, the main drawback of this method is that such simple partitioning is not scalable.
Overly aggressive point cloud partitioning, such as splitting a point cloud into one point per chunk, can lead to a significant accuracy drop.

To address this, for global-dependent operations, we group and operate multiple chunks together similar to stencil operations.
As \Fig{fig:split_lb} shows, we split the point cloud into 4 chunks and feed these chunks into a line buffer, $LB_1$.
Each time, we read one chunk, but the global-dependent operation does not start until a group of chunks (e.g., $1\times2$ chunks) has arrived.
In this case, the global-dependent operation reads $C_0$ and $C_1$ before the kNN search starts. 
For the next chunk of the kNN search, only $ C_2$ needs to be read, as $C_1$ was already read during the previous operation.

This is equivalent to partitioning the point cloud into $1\times4$ chunks and applying $1 \times 2$ stencil kernel with a stride of 1.
Although we use 1D stencil operation as an example, our technique can also be extended to 2D stencil operation. 
This way, we can achieve finer-grained pipelining while avoiding accuracy drops due to overly aggressive partitioning.

\paragraph{Pipelining.} 
Another advantage of compulsory splitting is that it enables finer-grained pipelining. 
\Fig{fig:split_example} shows a simplified pipeline in PointNet++~\cite{qi2017pointnet++} that includes a range search operation, $\cR$, sandwiched between a scaling operation, $\cS$, and multilayer perception, $\cM$. 
By default, $\cS$ and $\cR$ are executed sequentially due to the global data dependency of $\cR$. 
Only the executions of $\cR$ and $\cM$ can be overlapped.

As shown in \Fig{fig:split_example}, both naive splitting and compulsory splitting can achieve finer-grained pipelining.
The key difference between these two techniques is that \textit{naive splitting isolates each chunk during the execution, its operation accuracy is more susceptible to the number of chunks}.
Our technique mimics a ``coarse-grained'' stencil operation and groups adjacent chunks, e.g., $C_0$ and $C_1$ in the dashed red box, in global-dependent operations.
Since chunks like $C_0$+$C_1$ and $C_1$+$C_2$ can be processed independently, we can exploit pipelining between them. 
Doing so, we can retain some data dependency across neighboring chunks and preserve the accuracy of global-dependent operations while still retaining finer-grained pipelining.
\Sect{sec:eval:sens} shows the impact of our technique on the application performance and accuracy. 

It is worth noting that the example in \Fig{fig:split_example} simplifies the execution of $\cR$ by ignoring its non-deterministic behavior. 
In reality, each segment of $\cR$ execution (e.g., $\cR_0$) varies.
In \Sect{sec:algo:dt}, we address this non-determinism.

\paragraph{Split for Sorting.} 
Compulsory splitting can be applied to sorting as well. 
Since the spatial partitioning of a point cloud already determines the sorting order of different split chunks, sorting within each chunk naturally establishes the overall sorting order, similar to hierarchical sorting~\cite{murtagh2012algorithms}.

\paragraph{How to Split.} 
In practice, we split point clouds based on their harvesting processes.
For point clouds generated from CAD models, we use spatial partitioning to divide point clouds into spatially even chunks.
For LiDAR-generated point clouds, point cloud data are naturally serialized when produced by the LiDAR sensor; the order in which the points are produced exhibits locality because LiDAR scans points sequentially. 
Thus, we partition the LiDAR point clouds into even chunks (e.g., Points 1 to $N$ in chunk 1, and $N+1$ to $2N$ in chunk 2, etc.). 
$N$ is empirically chosen for performance.

\paragraph{When to Split.} 
Compulsory splitting enables finer-grained pipelining, a natural question is when to determine the number of chunks for global-dependent operations.
Our strategy is to apply uniform partitioning on the input point cloud \textit{offline}, so that all global-dependent operations in the pipeline adhere to this uniform splitting scheme. 
The rationale is that all global operations compute and establish spatial relationships between points in Cartesian coordinates, partitioning the original point cloud also aligns with this principle.
More fine-grained splitting strategies are left for future work.

\subsection{Deterministic Termination}
\label{sec:algo:dt}

Another challenge when applying the line buffer technique to point cloud applications is that some global-dependent operations, such as kNN search, are input-dependent.
Input-dependent operations lead to two issues.
First, the delay of a global-dependent operation varies, making it impossible to determine the required line buffer size between itself and the subsequent stage offline.
Second, operations like range and kNN search involve traversing tree-like data structures, leading to irregular memory accesses.

\paragraph{Variable Delay.} To address the variable delay of global-dependent operations, we observe that, for some input data, selectively skipping some steps can align execution delays while yielding results that closely approximate those of the canonical algorithms.
Based on the observation, our idea is to convert those ``delay-non-deterministic'' operations into deterministic ones. 
Specifically, our idea imposes a termination ``deadline'' on each non-deterministic operation. 
For example, in a kNN search, the deadline would be the number of tree traversal steps.
By setting a fixed deadline for each operation, all processes are completed within a fixed timeframe.
This way, we can determine the line buffer size offline.
We name our technique, \textit{deterministic termination}. 




\begin{figure}[t]
\centering
\includegraphics[width=0.9\columnwidth]{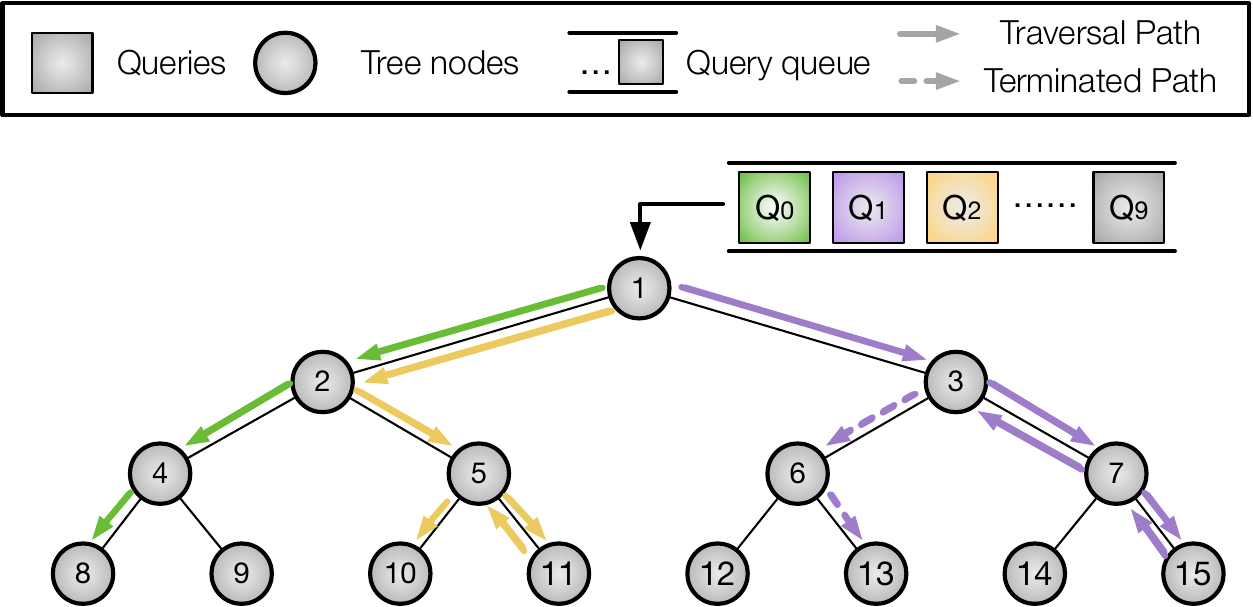}
\caption{An example of deterministic termination on kd-tree traversal. Here, we define the termination threshold to be 5. Different colors correspond to different query traversals. 
The dashed part is early stopped by deterministic termination. 
}
\label{fig:deterministic_stop}
\end{figure}

\Fig{fig:deterministic_stop} illustrates how we apply deterministic termination to a kd-tree search. 
In canonical kd-tree search, each query traverses the tree to find its nearest neighbor.
For instance, query $Q_0$ traverses 3 steps and finds its nearest neighbor, \circled{white}{8}, while $Q_1$ traverses $7$ steps to find \circled{white}{13}.
For each query, the termination criterion, i.e., the number of tree traversal steps, is dependent on the specific input.

With deterministic termination, we set a fixed deadline for every query. 
For example, in \Fig{fig:deterministic_stop}, each query’s traversal is capped at 5 steps. 
In this case, both queries $Q_0$ and $Q_2$ can be completed within this deadline. 
However, query $Q_1$, which originally required 7 steps to complete, is forced to be terminated at step 5 and returns the search results up to that point. 
This approach ensures uniform processing times across queries, which allows the minimal line buffer size to be predetermined and fixed throughout execution.

Another question to this approach is to determine an appropriate termination deadline for each operation. 
In our experiment, we establish these termination deadlines based on offline profiling. 
Our results show that setting a termination deadline for point cloud operations does not affect the results after an integrated co-training procedure described in \Sect{sec:algo:train}.
More exhaustive approaches to determine the deadlines are left for future work.

\paragraph{Irregular Memory Access.} 
The second issue with input-dependent operations is that they often involve irregular data access, such as kd-tree traversal. Due to the input-dependent nature of memory access patterns, any offline data layout optimization cannot fully eliminate on-chip bank conflicts. 
Although a few techniques in the literature address on-chip bank conflicts~\cite{khan2014padding, afshani2015sorting}, we address bank conflicts by adopting a simple strategy, the bank conflict elision technique from Crescent~\cite{feng2022crescent}. 
Whenever there is a bank conflict, only one of the requests is allowed to proceed.
The remaining requests bypass the rest of the data structure beneath the conflict node. 
Hence, we claim no contribution to this end. 

\begin{figure}[t]
\centering
\includegraphics[width=\columnwidth]{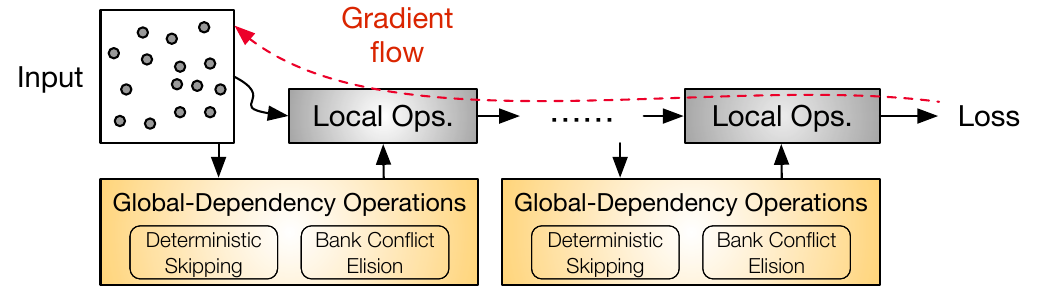}
\caption{Our training procedure integrates our proposed techniques. Although our techniques are not differentiable, the back-propagation does not flow through our optimizations.}
\label{fig:cotrain}
\end{figure}

\subsection{Integrated Co-Training}
\label{sec:algo:train}

Coupled with our two algorithmic optimizations, we also proposed a co-training procedure to mitigate the potential accuracy loss from these optimizations, as shown in \Fig{fig:cotrain}. 
Specifically, we integrate both algorithmic behaviors into the training process.
However, a challenge in this integration is that simulating compulsory splitting and deterministic termination are not differentiable. 
Our observation is that, in point cloud applications, global-dependent operations are used to establish relationships between points rather than directly manipulate point values. 
Thus, we can simulate these optimizations during training without concern for their differentiability. 
The back-propagation gradient flows only through the local-dependent operations, not the global-dependent operations.
\Sect{sec:eval:acc} presents an accuracy comparison with and without our co-training procedure.

\section{Line Buffer Optimization}
\label{sec:opti}

Our techniques in \Sect{sec:algo} relax the global dependencies in point cloud algorithms and convert non-deterministic point cloud pipelines into deterministic ones.
This conversion enables fine-grained pipelining across stages.
However, given a converted point cloud pipeline, the remaining challenge is to determine the minimal on-chip buffer size while ensuring that intermediate off-chip traffic is eliminated.

\Sect{sec:opti:overview} describes the general idea and algorithmic requirements behind our line buffer optimization. 
\Sect{sec:opti:eqn} describes how to formulate the optimization into an ILP problem.

\subsection{General Idea}
\label{sec:opti:overview}

\paragraph{Goal.} Similar to prior works on image processing~\cite{chi2018soda, ragan2013halide, hegarty2014darkroom}, 
we define point cloud pipelines as dataflow graphs.
Our optimization asks users to define the data dependency and communication pattern of each stage in the dataflow graphs. 
The optimization goal is to achieve the highest throughput while minimizing the overall on-chip buffer size.

\paragraph{Requirements.} To achieve this goal, our framework must meet three requirements:
\begin{itemize}
    \item First, any input elements read by a consumer stage must be produced by a preceding producer stage and be available in the line buffer.
    \item Second, once data is read from off-chip memory, it will not be written back to the off-chip memory until all computations are complete.
    \item Third, the pipeline must ensure that there are no on-chip memory stalls during the entire execution, simplifying the control logic.
\end{itemize}


With these requirements, we describe our hierarchical optimization. 
Recall, our algorithm splits the point cloud into multiple chunks. 
For example, in \Fig{fig:split_example}, $C_0$ and $C_1$ are one chunk in compulsory splitting.
For a single-chunk processing pipeline, our framework ensures the highest throughput while minimizing the on-chip buffer size. 
For each producer-consumer pair, the consumer is delayed by a specific number of cycles to ensure that, once it starts computing, all data dependencies are resolved, allowing it never to stall.

Once the line buffer sizes for a single-chunk pipeline are determined, we extend this result to multi-chunk pipelines with the same performance target.

\subsection{Optimization Formulation}
\label{sec:opti:eqn}
Here, we first describe the ILP formulation of line buffer optimization for a single chunk of point cloud. 

\paragraph{Objective.} The goal of our optimization is to minimize the overall line buffer size as shown in \Eqn{eqn:obj}:
\begin{align}
    \underset{\Theta}{\operatorname{argmin}}\ LB(\Theta) & = \sum_{i = 0}^{N-1}LB_i(\Theta), \nonumber \\
    \text{where}\ \Theta & = \{t_{i, s}\},\ i\in [0, 1, ..., N-1].
\label{eqn:obj}
\end{align}

In \Eqn{eqn:obj}, $LB_i(\Theta)$ is the size of the $i$th line buffer between stage $S_{i}$ and stage $S_{i+1}$, where $\Theta$ is the overall scheduling, i.e., the starting cycle $t_{s, i}$ of each stage $S_i$.

Next, we explain the two primary categories of constraints within our formulation: \textit{buffer size constraints} and \textit{data dependency constraints}.

\paragraph{Buffer Size Constraints.} For each line buffer $LB_i$, we guarantee that any line buffer $LB_i$ is large enough to accommodate all the elements during its execution. 
Mathematically, each line buffer size can be expressed as:
\begin{equation}
    LB_i(\Theta) = \max_{t\in [t_{w, i},\ t_{e, i}]} \{\#_{write, i, t} - \#_{overwrite, i, t}\},
\label{eqn:lb}
\end{equation}
where $t_{w, i}$ denotes the timestamp when stage $S_i$ starts writing to the line buffer $LB_i$, and $t_{e, i}$ denotes the timestamp at the end of writing to $LB_i$. 
$\#_{write, i, t}$ and $\#_{overwrite, i, t}$ denote the number of element written and overwritten to $LB_i$ at timestamp $t$, respectively. 
Therefore, $LB_i$ needs to accomondate $(\#_{write, i, t} - \#_{overwrite, i, t})$ number of elements at any timestamp between $t_{w, i}$ and $t_{e, i}$.
The constraint scope is narrowed between $t_{w, i}$ and $t_{e, i}$, because before $t_{w, i}$, $LB_i$ contains no elements, and after $t_{e, i}$, there is no more writes to $LB_i$.

Furthermore, $\#_{write, i, t}$ can be expressed as:
\begin{align}
    \#_{write, i, t} & = (t-t_{w,i}) \cdot \tau_{out,i}  = (t-(t_{s,i}+\Delta t_{stage, i})) \cdot \tau_{out,i},
\end{align}
where $\#_{write, i, t}$ represents the product of output throughput, $\tau_{out, i}$, of stage $S_i$ and time elapsed from $t_{w,i}$, which is further defined as the sum of $S_i$'s starting cycle $t_{s,i}$ 
and $\Delta t_{stage, i}$, which is the number of stages of $S_i$.
$\tau_{out, i}$ can be defined by $\rho_{out, i}$/$f_{out, i}$.
$\rho_{out, i}$ is the output shape of $S_i$ and $f_{out, i}$ is the output frequency of $S_i$.
Defining the output shape and frequency of $S_i$ is required to determine the data dependency.

Similarly, $\#_{overwrite, i, t}$ can be expressed as:
\begin{align}
    \#_{overwrite, i, t} & = \max\{0, t-t_{o, i}\} \cdot \tau_{in, i+1},
\end{align}
where $\tau_{in, i+1}$ is the input throughput of stage $S_{i+1}$. 
Whereas, $t_{o, i}$, the time at which elements start to be overwritten in $LB_i$ is a function of the data dependency of the given operation:
\begin{align}
    t_{o, i} & = \max_{c\in \cC_{i}} \{\cF(c)\}, \\
    \text{where}\ & \cF(c) = \begin{cases}
       t_{s,c},\ \text{if}\ S_i\ \text{is}\ \text{local}\ \text{dependent}, \nonumber \\
       t_{e,c},\ \text{otherwise}.                          \nonumber \\
    \end{cases}                                 
\end{align}

Here, $\cC_i$ is a collection of consumers of stage $S_i$. If the consumer is an operation with local data dependency, $t_{o, i}$ only requires to be greater than the starting of the consumer, $t_{s,c}$, otherwise, $t_{o, i}$ needs to be greater than the end of the consumer, $t_{e, c}$, which will be explained in \Eqn{eqn:global_dep}. 

\paragraph{Data Dependency Constraints.} In addition to the line buffer constraints, the operations also have inherent constraints determined by their data dependencies. 
These constraints can be categorized into those for local-dependent operations and those for global-dependent operations.

For local-dependent operations, we need to ensure that any element required by a later stage has already been written to the line buffer beforehand. 
Therefore, for each producer-consumer pair, the data dependency constraint can be generalized as follows:
\begin{align}
    \forall t \in [t_{s, i+1}, t_{e, i+1}],\ & (t-(t_{s, i} + \Delta t_{stage, i})) \cdot \tau_{out, i} \nonumber \\ 
    & \geq (t-(t_{s, i+1} + \Delta t_{stage, i+1})) \cdot \tau_{in, i+1},
\label{eqn:local_dep}
\end{align}
where $\tau_{out, i}$ and $\tau_{in, i+1}$ represent the output throughput of $S_i$ and the input throughput of $S_{i+1}$, respectively. 
\Eqn{eqn:local_dep} ensures that the total number of elements required by the consumer $S_{i+1}$ is always less than or equal to the number of elements produced by the producer $S_i$.


Specifically, in terms of stencil operations, 
it is essential to ensure that all elements required by the sliding window are available when needed:
\begin{align}
    \tau_{out, i} = \frac{\rho_{out, i}}{f_{out, i}},\ \tau_{in, i+1} = \frac{\rho_{in, i+1}}{\beta_{i+1} \cdot f_{in, i+1}}, \nonumber
\end{align}
where $\rho_{in, i+1}$ and $f_{in, i+1}$ are the input shape and frequency of $S_{i+1}$, respectively.
$\beta_{i+1}$ represents the input reuse factor, indicating the number of times each input element is reused at stage $S_{i+1}$.

For reduction operations, where a group of input elements contributes to a single output, $\tau_{out, i}$ and $\tau_{in, i+1}$ can be expressed as:
\begin{align}
    \tau_{out, i} = \frac{\rho_{out, i}}{f_{out, i}},\ \tau_{in, i+1} = \frac{\rho_{in, i+1}}{f_{in, i+1}}, \nonumber
\end{align}

For global-dependent operations, where potentially all elements need to be accessed. The data dependency constraint in \Eqn{eqn:local_dep} needs to be modified:
\begin{align}
    t_{e, c} \leq t_{s, i+1} \Rightarrow t_{s, i} + \Delta t_{stage, i} + \frac{W_i}{\tau_{out, i}} \leq t_{s, i+1},
\label{eqn:global_dep}
\end{align}
where $W_i$ is the total number of output elements produced by stage $S_i$.
This can be inferred from the dataflow graph.
\Eqn{eqn:global_dep} guarantees that ensures that all elements produced by stage $S_i$ are available before stage $S_{i+1}$ starts.

\paragraph{Constraint Pruning.} One issue with this optimization formulation is that the line buffer constraint in \Eqn{eqn:lb} and the local-dependency constraint in \Eqn{eqn:local_dep} impose a large number of constraints and variables.
For example, PointNet++~\cite{qi2017pointnet++} requires >100K constraints, making solving the optimization infeasible.
To address this, we observe that the equations for $\#_{write, i, t}$ and $\#_{overwrite, i, t}$ are monotonically increasing. 
Therefore, the constraints in \Eqn{eqn:lb} can be simplified into two, one is when $t$ is $t_{w, i}$ and the other is when $t$ is $t_{e, i}$:
\begin{align}
        LB_i (\Theta) & =  \max \bigl\{ ((t_{o,i}-(t_{s,i}+\Delta t_{stage, i})) \cdot \tau_{in, i+1}) , \nonumber \\
                 & ((t_{e, i} - (t_{s,i}+\Delta t_{stage, i})) \cdot \tau_{out, i}-(t_{e, i} - t_{o, i}) \cdot \tau_{in, i+1})\bigl\}, \nonumber \\
                 & \text{where} \ t_{e, i} = t_{s, i} + \Delta t_{stage, i} + \frac{W_i}{\tau_{out, i}}.
\end{align}


\begin{figure}[t]
\centering
\includegraphics[width=\columnwidth]{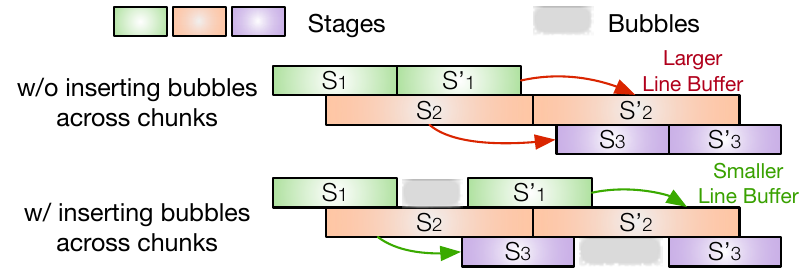}
\caption{An example of inserting bubbles across processing different chunks. Inserting bubbles reduces the line buffer sizes while retaining the same performance.}
\label{fig:bubble}
\end{figure}

\paragraph{Multi-Chunk Optimization.} 
Extending to multi-chunk pipelines, our framework retains the same line buffer sizes from the single-chunk optimization and keeps the same performance target. 
We observe that collapsing the execution of multiple chunks into a continuous process results in a larger line buffer without providing any performance gains, as shown in the upper part of \Fig{fig:bubble}. 
To ensure the same on-chip buffer can accommodate all intermediate data across chunks, we introduce bubbles at the start of stages that have shorter execution times, e.g., $S_1$ and $S_3$. 
These bubbles help manage dataflow and prevent buffer overflows, all while avoiding the additional on-chip resources.

\section{Programming Interface}
\label{sec:framework}

While our optimization described in \Sect{sec:opti} can determine the optimal line buffer sizes given a point cloud pipeline, it requires data dependencies and communication patterns between stages, as outlined in \Tbl{tab:variables}. 
However, automatically inferring these parameters is challenging.
Thus, we provide a programming interface that allows programmers to explicitly define these parameters.

Although there are existing line buffer optimization interfaces for image signal processing~\cite{chi2018soda, ujjainkar2023imagen, hegarty2014darkroom, whatmough2019fixynn}, 
these interfaces are designed primarily for stencil operations and are insufficient to express the data dependencies and communication patterns specific to point cloud applications, as shown in \Sect{sec:bg}.
To accommodate these diverse operations, we propose an interface that allows users to easily define the dataflow of their point cloud applications.




\begin{table} 
\caption{List of symbols used in the input of our framework, along with the mapping between the variables in the ILP optimization and the input parameters in our interface.}
\resizebox{\columnwidth}{!}{
\renewcommand*{\arraystretch}{1}
\renewcommand*{\tabcolsep}{5pt}
\begin{tabular}{ ccl } 
\toprule[0.15em]
\textbf{\specialcell{Symbols}} & \textbf{\specialcell{Parameter\\Names}} & \textbf{Definitions} \\ 
\midrule[0.10em]
$\rho_{in,i}$        & \texttt{i\_shape} & Input shape of stage $i$ \\
$f_{in,i}$           & \texttt{i\_freq}  & Input read frequency of stage $i$ \\
$\beta_{i}$          & \texttt{reuse}    & Input reuse pattern of stage $i$ \\
$\Delta t_{stage,i}$ & \texttt{stage}    & Number of pipelining stages of stage $i$ \\
$\rho_{in,i}$        & \texttt{o\_shape} & Output shape of stage $i$ \\
$f_{in,i}$           & \texttt{o\_freq}  & Output written frequency of stage $i$ \\
\bottomrule[0.15em]
\end{tabular}
}
\label{tab:variables}
\end{table}

\begin{figure}[t]
\centering
\includegraphics[width=\columnwidth]{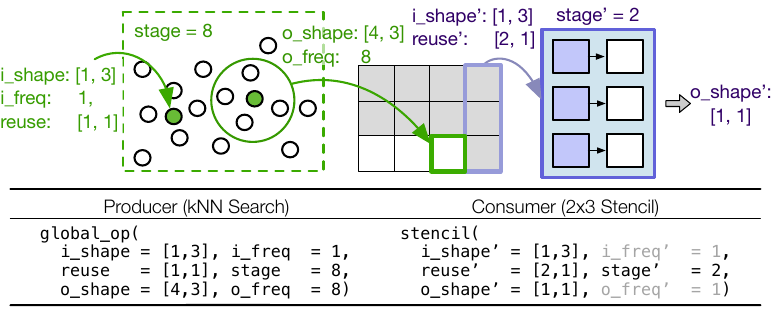}
\caption{An example of defining the dataflow graph using our framework interface. Our interface allows users to define the data dependencies and communication patterns. The greyish parameters do not need to be defined and can be inferred from the computation pattern.}
\label{fig:code}
\end{figure}

\paragraph{Interface.} 
Due to the diversity in point cloud operations in \Sect{sec:bg:pc}, it is hard to exhaust all the operations.
Thus, our interface provides a set of abstract operations that allow users to define the data dependencies and communication patterns of point cloud pipelines, without specifying the actual computation of each operation. 
\Fig{fig:code} illustrates an example of describing a pipeline including a kNN search operation followed by a stencil operation. 

For each operation, our interface uses two key parameters: \texttt{i\_shape} and \texttt{i\_freq}, which define the input pattern. 
\texttt{i\_shape} specifies the input shape, while \texttt{i\_freq} defines how frequently this operation reads input data with the specified input shape. 
The \texttt{i\_shape} is a tuple, such as $[x, y]$, where `x' represents the number of points and `y' is the number of attributes per point.
Similarly, \texttt{o\_shape} and \texttt{o\_freq} describe the output shape and frequency.

To capture data dependencies, the \texttt{reuse} parameter is used to define input reuse patterns, and \texttt{stage} indicates the number of pipeline stages of the operation.
The \texttt{reuse} parameter is also a tuple that specifies how many times each element is reused at each dimension.

Specifically, in \Fig{fig:code}, the 8-stage kNN search reads $1 \times 3$ inputs per cycle and produces $4 \times 3$ outputs every 8 cycles, with no input reuse. 
On the other hand, the 2-stage stencil operation performs a $2\times3$ stencil, reading $1\times3$ elements and producing $1\times1$ outputs per cycle. 
Each element of a 2x3 stencil operation will be used in the first dimension as defined by \texttt{reuse} parameter.
The stencil operation's input and output frequency are implicitly defined as 1.




\paragraph{Supported Operations.} 
Our framework includes a list of abstract operations listed below, allowing users to accommodate their applications. 
Some parameters are not specified as they can be inferred from the computation patterns. 
\begin{lstlisting}[language=Python, caption=Programming interface., numbers=none]
# stencil operation
stencil(i_shape, o_shape, stage, reuse)
# reduction operation
reduction(i_shape, o_shape, stage, o_freq)
# global operation
global_op(i_shape, o_shape, i_freq, o_freq, reuse, stage)
\end{lstlisting}
\vspace{-5pt}

\section{Experimental Setup}
\label{sec:exp}

\paragraph{Experimental Methodology.} We evaluate the generated architectures from our framework with the existing architectures. 
Specifically, we develop RTL implementation for the primary IP blocks and synthesize these IP blocks using Synposys synthesis and Cadence layout tools in TSMC 16nm FinFET technology. 
We use Google's optimization library "OR-Tools" to solve the ILP optimization in \Sect{sec:opti} and finalize each SRAM size.
The SRAMs are generated using the Arm Artisan memory compiler. 
Power is estimated using Synopsys PrimeTimePX with annotated switching activities.
The DRAM parameters are modeled after Micron 16 Gb LPDDR3-1600 (4 channels) according to its datasheet~\cite{micronlpddr3}.
Lastly, we build a cycle-level simulator of the architecture with the latency of each component parameterized from the post-synthesis results of the ASIC designs.

\paragraph{Applications.} \Tbl{tab:eval_app} lists the applications used for evaluation across four domains: classification, segmentation, registration, and neural rendering.

For classification, we use the PointNet++(c)\cite{qi2017pointnet++} algorithm and evaluate on two datasets, ModelNet10\cite{wu20153d} and ModelNet40~\cite{wu20153d}, using overall accuracy as the metric. 
For segmentation, we use PointNet++(s) with the ShapeNet dataset~\cite{chang2015shapenet}, using the standard mean Intersection-over-Union (mIoU) as our accuracy metric.
For registration, we use the A-LOAM algorithm~\cite{zhang2014loam} on the KITTI dataset~\cite{geiger2013vision}, with end-to-end translation and rotation errors as the accuracy metrics. 
For rendering applications, we utilize 3D Gaussian splatting (3DGS)~\cite{kerbl20233d} as the rendering pipeline and assess performance on the Tank\&Temple~\cite{knapitsch2017tanks} and DeepBlending~\cite{hedman2018deep} dataset with the standard Peak Signal-to-Noise Ratio (PSNR) metric.

\begin{table} 
\caption{Evaluation benchmarks.}
\resizebox{\columnwidth}{!}{
\renewcommand*{\arraystretch}{1}
\renewcommand*{\tabcolsep}{5pt}
\begin{tabular}{ ccccc } 
\toprule[0.15em]
\textbf{\specialcell{Application\\Domains}} & \textbf{\specialcell{Algorithm}} & \textbf{Dataset} & \textbf{\specialcell{Hardware\\Baselines}} & \textbf{\specialcell{Global\\Dependencies}} \\ 
\midrule[0.05em]
\specialcell{Classification} &  PointNet++ (c) & \specialcell{ModelNet10,\\ModelNet40} & \specialcell{PointAcc,\\Mesorasi} & \specialcell{Range\\Search}\\
\midrule[0.05em]
\specialcell{Segmentation} &  
PointNet++ (s) & ShapeNet & \specialcell{PointAcc,\\Mesorasi} & \specialcell{Range\\Search}\\
\midrule[0.05em]
\specialcell{Registration} &  \specialcell{A-LOAM} & KITTI & \specialcell{QuickNN,\\Tigris} & \specialcell{kNN\\Search}\\
\midrule[0.05em]
\specialcell{Neural\\Rendering} &  \specialcell{3DGS} & \specialcell{Tank\&Temple,\\DeepBlending} & \specialcell{GScore} & Sorting\\
\bottomrule[0.15em]
\end{tabular}
}
\label{tab:eval_app}
\end{table}

\paragraph{Hardware Baselines.} 
To date, there is no single accelerator capable of supporting all the point cloud applications we evaluated. 
Here, we compare five existing point cloud architectures with the design obtained from our framework. 
For a fair comparison, we configure the hardware baselines with the same amount of processing elements (PEs) and comparable on-chip buffers. 
For classification and segmentation tasks, we compare two deep learning accelerators, PointAcc~\cite{lin2021pointacc} and Mesorasi~\cite{feng2020mesorasi}. 
For registration, we compare two point cloud kNN accelerators, QuickNN~\cite{pinkham2020quicknn} and Tigris~\cite{xu2019tigris}. 
Lastly, we compare against GScore~\cite{lee2024gscore} on neural rendering.

\paragraph{Variants.} We evaluate three variants in our paper:
\begin{itemize}
\item \mode{Base}: This is a variant with no compulsory splitting or deterministic termination.
\item \mode{Base+\$}: This variant is similar to \mode{Base}, but it replaces line buffers with a fully-associative cache.
\item \mode{CS}: This variant only uses compulsory splitting without deterministic termination.
\item \mode{CS+DT}: This is the full-fledged version with both compulsory splitting and deterministic termination.
\end{itemize}

\section{Evaluation}
\label{sec:eval}

\Sect{sec:eval:acc} shows that our two optimizations bring marginal accuracy loss. \Sect{sec:eval:perf} shows that, under the same throughput, our techniques reduce on-chip buffer size and energy consumption compared to the hardware baseline.
\Sect{sec:eval:prior} compares against prior state-of-the-art accelerators and \Sect{sec:eval:sens} shows sensitivity studies. 

\subsection{Accuracy}
\label{sec:eval:acc}

We evaluate our two techniques, \textit{compulsory splitting} (\mode{CS}) and \textit{deterministic termination} (\mode{DT}), across four domains to show the wide applicability of our optimizations.
We describe our setting for each task first, and unless otherwise noted, the same setting is used throughout the evaluation.

\begin{figure}[t]
\centering
\begin{minipage}[t]{0.48\columnwidth}
  \centering
  \includegraphics[width=\columnwidth]{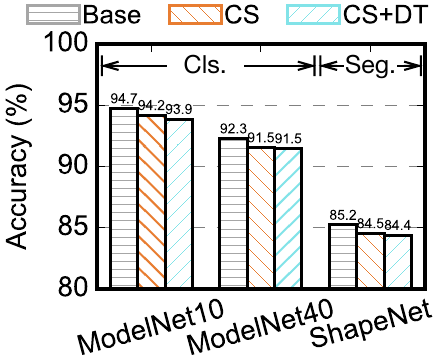}
  \caption{Accuracy on classification and segmentation.}
  \label{fig:cls_seg_acc}
\end{minipage}
\hspace{2pt}
\begin{minipage}[t]{0.48\columnwidth}
  \centering
  \includegraphics[width=\columnwidth]{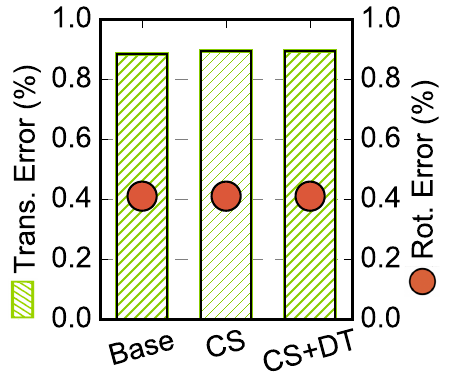}
  \caption{Accuracy comparison on registration.}
  \label{fig:reg_acc}
\end{minipage}
\end{figure}

\paragraph{Classification and Segmentation.} 
\Fig{fig:cls_seg_acc} shows the accuracy comparison between the baselines and our techniques on two tasks: classification and segmentation. 
We split each point cloud into $3\times3\times1$ chunks with a kernel size of $2\times2$, which is equivalent to partitioning the point cloud into 4 chunks. 
We set the termination deadline to approximately 25\% of the steps compared to a full kd-tree traversal. 
Applying \mode{CS} alone results in an average accuracy loss of only 0.6\%.
By applying both techniques, \mode{CS+DT} retains accuracy loss with less than 1\% (0.8\% on average).

\paragraph{Registration.}
\Fig{fig:reg_acc} shows the comparison of translational and rotational errors between our techniques and the baseline algorithm, A-LOAM. 
In this task, it is equivalent to partitioning the point cloud into 4 chunks. The termination deadline is also 25\% of the full kd-tree traversal. 
In \Fig{fig:reg_acc}, our approach introduces marginal accuracy loss compared to the baseline algorithm. On average, we introduce only 0.01\% additional translational error and no rotational error.

\begin{figure}[t]
\centering
\begin{minipage}[t]{0.48\columnwidth}
  \centering
  \includegraphics[width=\columnwidth]{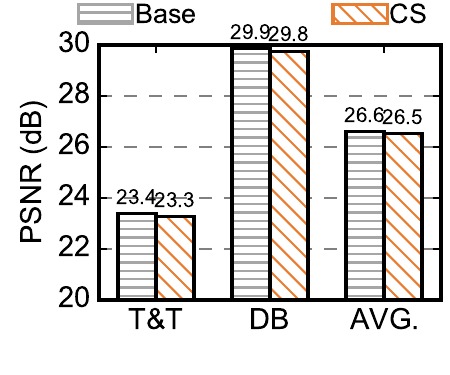}
  \caption{Accuracy comparison on neural rendering.}
  \label{fig:render_acc}
\end{minipage}
\hspace{2pt}
\begin{minipage}[t]{0.48\columnwidth}
  \centering
  \includegraphics[width=\columnwidth]{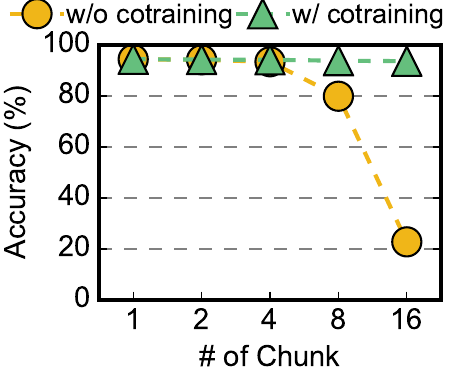}
  \caption{Accuracy with and without co-training.}
  \label{fig:cotrain_res}
\end{minipage}
\end{figure}

\begin{figure}
\centering
\subfloat[On-chip buffer reduction.]{
  \includegraphics[width=0.46\columnwidth]{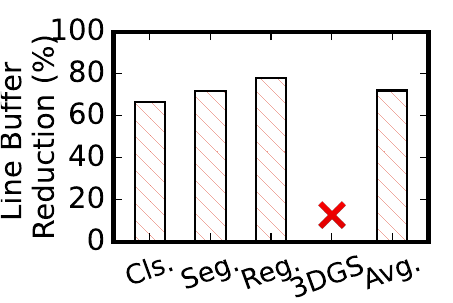}
  \label{fig:buf_reduce}
}
\subfloat[Normalized energy.]{ 
  \includegraphics[width=0.46\columnwidth]{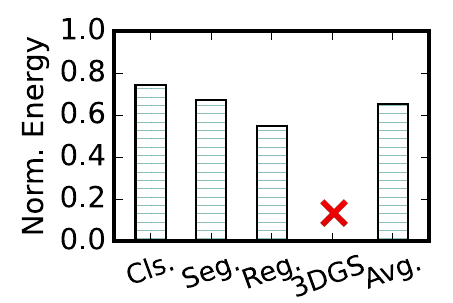}
  \label{fig:energy}
}
\caption{On-chip buffer reduction and energy savings by applying compulsory splitting and deterministic termination.} 
\label{fig:splitPerformance}
\end{figure}

\paragraph{Rendering.}
\Fig{fig:render_acc} compares the rendering quality between our techniques and the baseline algorithm, 3DGS. 
Since 3DGS rendering typically requires high storage (>1GB), we partition the entire point cloud into $80 \times 60 \times 75$ chunks with a stride of $1 \times 1 \times 1$.
Because no non-deterministic operations in 3DGS, \mode{DT} does not apply to 3DGS. 
We evaluate our approach against two datasets and find that, on average, we achieve similar accuracy compared to the baseline with a negligible loss of 0.1 dB in PSNR.

\paragraph{Integrated Co-Training.}
\Fig{fig:cotrain_res} shows the accuracy comparison with and without our co-training on classification. 
Other tasks follow a similar trend. 
In \Fig{fig:cotrain_res}, models without co-training can maintain high accuracy when the number of chunks is low. 
However, as the number of chunks increases, the accuracy drops rapidly. 
In contrast, our co-training integrates algorithmic behavior, making the trained model robust and capable of retaining high accuracy even when the number of chunks is high.
Note that, the co-training overhead is 3.1$\times$ slower compared to the baseline.
This overhead is primarily due to inefficient CPU kernels simulating \mode{DT} behaviors during training. 
More efficient GPU implementations are left for future work.

\subsection{Buffer Reduction and Energy Savings}
\label{sec:eval:perf}

This section shows that our techniques can effectively reduce the line buffer size and save energy. 
Note that, since our line-buffer optimization maintains the same target throughput, thus, our techniques do not improve performance.
Both our hardware and the baseline use the same amount of compute units, the only difference is the buffer size.

\Fig{fig:buf_reduce} shows the line buffer reductions compared to the baselines.
On average, we reduce the line buffer size by 72\%.
The 3DGS result is missing as the baseline requires an infeasibly large on-chip buffer ($>1$~GB), which our toolchain could not synthesize.
\Fig{fig:energy} presents the normalized energy consumption relative to the baselines.
On average, we achieve a 40.5\% reduction in energy.
This energy savings is entirely attributed to the reduction in SRAM sizes.


\subsection{Prior Work Comparison}
\label{sec:eval:prior}

\begin{figure}[t]
\centering
\subfloat[Classification.]{
  \includegraphics[width=0.46\columnwidth]{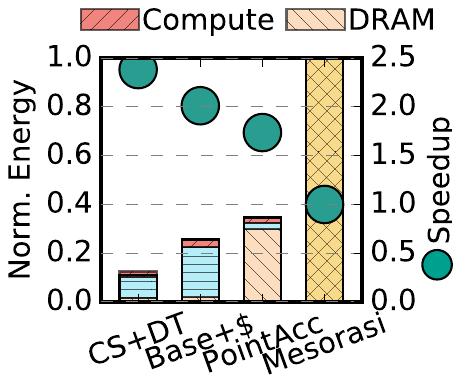}
  \label{fig:cls_comp}
}
\subfloat[Segmentation.]{
  \includegraphics[width=0.46\columnwidth]{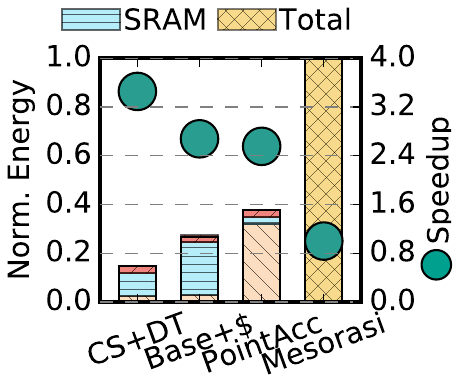}
  \label{fig:seg_comp}
}
\vspace{-5pt}
\subfloat[Registration.]{
  \includegraphics[width=0.46\columnwidth]{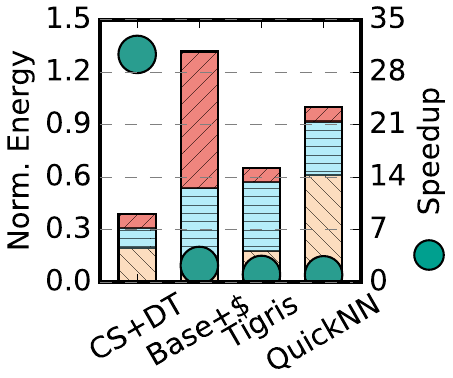}
  \label{fig:reg_comp}
}
\subfloat[3DGS.]{
  \includegraphics[width=0.46\columnwidth]{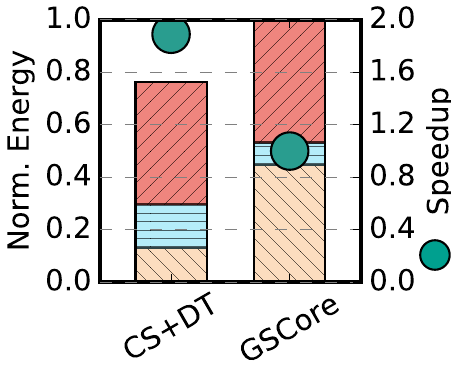}
  \label{fig:gaus_comp}
}
\caption{Overall performance and normalized energy comparison against prior accelerators. The legend is shared.}
\label{fig:prior_comp}
\end{figure}

In addition to comparing against the baselines, we also evaluate our design against prior point cloud accelerators. 
For a fair comparison, we ensure that both the prior works and our design have the same number of compute resources, i.e., 256 PEs, and a comparable overall on-chip buffer size.

\Fig{fig:cls_comp} shows the speedup and normalized energy between \mode{CS+DT} and two prior accelerators, \mode{PointAcc} and \mode{Mesorasi}, on classification. 
Both speedup and energy metrics are normalized against \mode{Mesorasi}. 
Our \mode{CS+DT} achieves 1.4$\times$ and 2.4$\times$ speedup compared to \mode{PointAcc} and \mode{Mesorasi}, respectively. 
We also compare against {\mode{Base+\$}} and achieve 1.2$\times$ speedup. 
Because cache misses would introduce frequent pipeline stalls and off-chip traffic.

In terms of energy savings, the primary contribution comes from reducing DRAM energy, as our streaming process eliminates intermediate off-chip traffic, resulting in a 94.4\% reduction in DRAM energy. 
Although our \mode{CS+DT} introduces additional SRAM energy due to a larger on-chip buffer (2.4~MB vs. 257~KB), we still achieve an overall energy reduction of 63.9\% compared to \mode{PointAcc}. 
Compared to \mode{Base+\$}, we achieve 57.2\% energy saving.
There is no energy breakdown of \mode{Mesorasi} because some of their energy numbers are measured directly from mobile GPU.
Segmentation shows a similar trend in \Fig{fig:seg_comp}.

\Fig{fig:reg_comp} shows the speedup and normalized energy against two prior accelerators, \mode{Tigris} and \mode{QuickNN}, on a registration task. 
Both speedup and energy metrics are normalized against \mode{QuickNN}. 
Our \mode{CS+DT} achieves 28.9$\times$ and 30.4$\times$ speedup over \mode{Tigris} and \mode{QuickNN}, respectively. 
The speedup primarily comes from a smaller search range enabled by compulsory splitting and deterministic termination, as kNN search is the main bottleneck in registration. 
\mode{CS+DT} also achieves 30.1\% and 60.8\% energy reduction over \mode{Tigris} and \mode{QuickNN} due to memory savings.
Compared to \mode{Base+\$}, \mode{CS+DT} achieves 13.1$\times$ speedup and 70.3\% energy reduction.

Lastly, we show the comparison against \mode{GSCore}. 
Values are normalized to \mode{GSCore}.
\Fig{fig:gaus_comp} shows that we achieves 1.9$\times$ speedup and 22.3\% energy reduction. 
The speedup comes from our streaming processing. 
The main energy contributor is reduced DRAM traffic.
\mode{Base+\$} does not apply to 3DGS due to the infeasible on-chip buffer described in \Sect{sec:eval:perf}.

\begin{figure}[t]
\centering
\subfloat[Classification.]{
  \includegraphics[width=0.46\columnwidth]{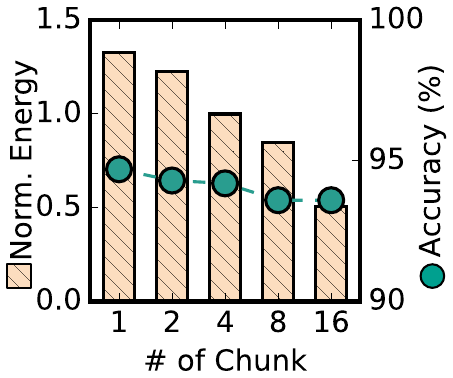}
  \label{fig:sens_cls}
}
\subfloat[Segmentation.]{
  \includegraphics[width=0.46\columnwidth]{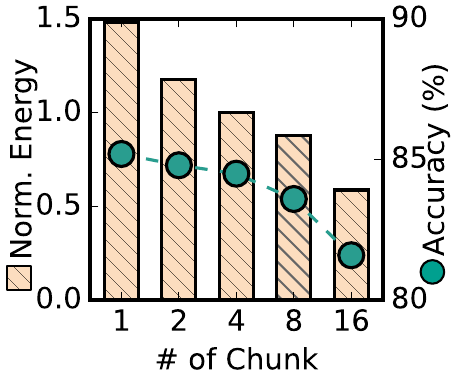}
  \label{fig:sens_seg}
}
\caption{Sensitivity of energy and accuracy to compulsory splitting, i.e., the number of split chunks.}
\label{fig:sens_sp}
\end{figure}

\subsection{Sensitivity Study}
\label{sec:eval:sens}

\paragraph{Compulsory Splitting.} 
\Fig{fig:sens_sp} illustrates how accuracy and energy change with different numbers of split chunks.
Here, we show the results of classification and segmentation.
We do not show the results of registration and rendering, because the number of chunks does not significantly affect the accuracy of registration and rendering.
In \Fig{fig:sens_sp}, the energy numbers are normalized against the scenario with 4 chunks.
As the number of chunks increases, classification accuracy shows a slight drop in \Fig{fig:sens_cls}, thanks to our co-training procedure. 
On the other hand, the normalized energy decreases significantly, as the on-chip buffer size reduces from 2.4MB at 4 chunks to 1.8MB at 16 chunks, leading to 49.6\% energy reduction. 
The energy reduction trend of the segmentation task is similar.
However, the segmentation accuracy drops more significantly as the number of chunks increases to 16, as shown in \Fig{fig:sens_seg}.
This shows that the sensitivity of accuracy to the number of chunks is task-specific.


\begin{figure}[t]
\centering
\subfloat[Classification.]{
  \includegraphics[width=0.46\columnwidth]{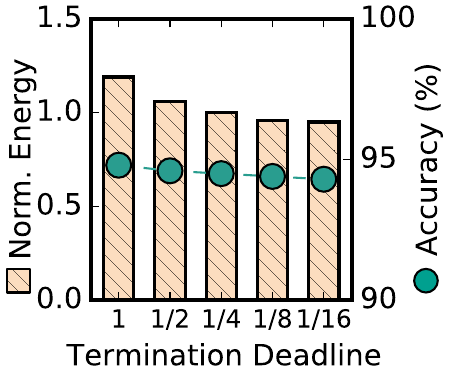}
  \label{fig:sens_dt_cls}
}
\subfloat[Segmentation.]{
  \includegraphics[width=0.46\columnwidth]{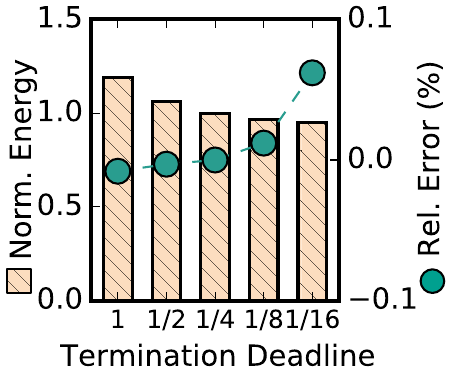}
  \label{fig:sens_dt_reg}
}
\caption{Sensitivity of energy and accuracy to deterministic termination.}
\label{fig:sens_dt}
\end{figure}

\paragraph{Deterministic Termination.}
{\Fig{fig:sens_dt}} presents the accuracy and energy sensitivity to the deterministic termination for classification ({\Fig{fig:sens_dt_cls}}) and registration ({\Fig{fig:sens_dt_reg}}). 
All results are normalized to the termination deadline to $\frac{1}{4}$ of a full kd-tree traversal.
The classification accuracy changes minimally while the registration accuracy declines as the termination deadline decreases.
However, our method still maintains comparable accuracy when the deadline is set to $\frac{1}{4}$.
Further shortening the deadline would lead to a significant accuracy drop.
Meanwhile, the overall energy consumption decreases with shorter deadlines. 
For instance, in {\Fig{fig:sens_dt_reg}}, the total energy reduces by 20\% when the deadline decreases from 1 to $\frac{1}{4}$ while the energy reduces merely by 5\% when the deadline decreases further to $\frac{1}{16}$.

\section{Related Work}

\paragraph{Point Cloud Acceleration.} 
Point cloud accelerators have gained significant attention in the hardware community recently, leading to many design innovations~\cite{zhang2021point, lin2021pointacc, feng2020mesorasi, xu2019tigris, ying2023edgepc, han2024bitnn, lyu2023flna, lian2024point, zheng2023tipu, feng2024cicero, feng2024potamoi}.
For instance, Mesorasi~\cite{feng2020mesorasi}, PointAcc~\cite{lin2021pointacc}, and Point-X~\cite{zhang2021point} focus on efficient accelerators for DNN-based algorithms, while Crescent~\cite{feng2022crescent} tackles irregular memory accesses in kNN search.
QuickNN~\cite{pinkham2020quicknn} and Tigris~\cite{xu2019tigris} utilize hierarchical kNN search to accelerate conventional point cloud pipelines like registration.
In contrast, \proj addresses the challenge of off-chip traffic by proposing a streaming processing approach, enabling fine-grained pipelining to accelerate point cloud applications.

\paragraph{Accelerator Design Framework.} 
Recently, several agile accelerator frameworks have emerged~\cite{durst2020type, lai2019heterocl, li2020heterohalide}. 
In the domain of image processing, frameworks such as Darkroom\cite{hegarty2014darkroom}, FixyNN\cite{whatmough2019fixynn}, and SODA\cite{chi2018soda} focus on optimizing accelerators through the use of line buffers.
Spatial~\cite{koeplinger2018spatial} introduces a domain-specific language for quick accelerator design, while DSAGEN~\cite{weng2020dsagen} expands this approach by automatically exploring the optimal accelerator design within a large architectural design space. 
However, these frameworks primarily target regular computations. 
\proj addresses the non-determinism in point cloud computation, leveraging line buffer techniques to optimize overall on-chip buffer size.
\section{Conclusions}
\label{sec:conc}

This paper presents \proj, a framework that transforms non-deterministic, global-dependent operations into deterministic ones and enables streaming processing by leveraging line buffer techniques. 
\proj achieves significant energy reduction and outperforms existing accelerators while maintaining comparable task accuracy. 
\section{Acknowledgements}

We thank anonymous reviewers from ASPLOS for their comments. The work is partially supported by the National Key R\&D Program of China under Grant 2022YFB4501400, NSFC grant (62402312 and 62222210).
This work is also supported by Shanghai Qi Zhi Institute Innovation Program SQZ202316.
Weikai Lin and Yuhao Zhu were not financially supported by the awards acknowledged by the other author(s) of this publication.

\bibliographystyle{plain}
\bibliography{references}

\end{document}